\documentclass[aps,preprint,showpacs,showkeys,floatfix,nofootinbib]{revtex4-1}

\usepackage{dcolumn}

\usepackage{extarrows}
\usepackage{graphicx}
\usepackage{graphicx,epstopdf}
\usepackage{dcolumn}
\usepackage{bm}

\newcommand{\nn}{\nonumber}

\begin{document}

\title[]{Second Post-Minkowskian Metric for a Moving Kerr Black Hole}
\author{Guansheng He, Chunhua Jiang, and Wenbin Lin}

\email{Email: wl@swjtu.edu.cn. }

\affiliation{
School of Physical Science and Technology, Southwest Jiaotong University, Chengdu 610031, China
}%

\date{\today}

\begin{abstract}
The harmonic metric for a moving Kerr black hole is presented in the second post-Minkowskian approximation. It is further demonstrated that the obtained metric is consistent with the Li\'{e}nard-Wiechert gravitational potential for a moving and spinning source with an arbitrary constant velocity. Based on the metric, we also give the post-Newtonian equations of motion for photon and massive test particle in the time-dependent gravitational field.
\end{abstract}


\maketitle

\section{Introduction} \label{sec:intro}
The time-dependent gravitational field motivated by moving source system is an attractive subject. Several groups have addressed the corrections due to the motion of source(s) to the propagation of a test particle theoretically. In order to study light propagation in the gravitational field of arbitrarily moving N-body system, Kopeikin and Sch\"{a}fer~\cite{KopeiSch1999} obtained their explicit first post-Minkowskian Li\'{e}nard-Wiechert gravitational potential. This retarded solution of field equations was validated via using a Lorentz boost to the equations of light propagation in the static field of source~\cite{Klioner2003}. The numerical simulations for light deflection by moving gravitational bodies were made in Ref.~\cite{KlionerPeip2003}, where the two authors considered the equations of light trajectory in both post-Newtonian and post-Minkowskian approximation, respectively. The new technique of integration in Ref.~\cite{KopeiSch1999} was also used to investigate the gravito-magnetism for light deflection in the weak field of moving and rotating bodies by Kopeikin and Mashhoon~\cite{KopeiMash2002}. According to the generalized idea of Fermat's principle, Sereno~\cite{Sereno2002b,Sereno2005b} considered the velocity-effect corrections of spinning source with low translational speed to gravitational lensing of light. Then, Sereno~\cite{Sereno2007} evaluated both the effects of peculiar motion and rotation of galaxy clusters on the properties of gravitational lensing. Bonvin~\cite{Bonvin2008} discussed the weak-field perturbation of peculiar motion of galaxies on the light propagation, and showed that it obviously affected the magnification and the relation between cosmic shear and convergence. Wucknitz and Sperhake~\cite{WuckSper2004} studied not only the light's gravitational deflection, but also a massive test particle's deflection, in the time-dependent field of a uniformly moving source.

On the other hand, the motion effects associated with observational experiments were also investigated extensively. The effects of the solar barycentric motion to post-Newtonian parameter $\gamma$ in Cassini spacecraft experiment were analyzed under Lorentz transformation (see, e.g. Refs.~\cite{KPSV2007,BAI2008,Kopeikin2009}). Kopeikin and Makarov~\cite{KopeiMak2007} calculated the planetary monopolar, dipolar and quadrupole deflection of light by moving source in solar system and discussed the possibilities for the actual observation of these effects. The authors of Refs.~\cite{ZscKlio2011,Zschocke2011} simplified the quadrupole formula given in Ref.~\cite{KopeiMak2007} to examine the general relativity in exploration of Gaia mission~\cite{Linet2008}. There are also increasing interests for measuring proper motion of stars and other deflection sources~\cite{CudHer1979,EckGen1997,Jarosz1998,Albroetal1999,KopeiMaka2006,Novatietal2010}.

When considering motion effects of gravitational source in theoretical calculations for high-order gravitational deflection and the equations of motion, usually we need to include the high-order effects produced by the source's mass. The high-order velocity effects can be calculated only when the metric of this kind of source is known.

In this paper we derive the second post-Minkowskian harmonic metric of a moving Kerr black hole with an arbitrary constant speed, and calculate the equations of motion of photon and massive test particle. We demonstrate the consistency of the solution obtained via Lorentz transformation with the retarded Li\'{e}nard-Wiechert potential~\cite{KopeiSch1999} for the first post-Minkowskian limit.

Throughout the units where $G=c=1$ are used. In addition, as done in Ref.~\cite{WuckSper2004}, we use the primed coordinates to distinguish the rest frame of the gravitational source $(t',~x',~y',~z')$ from that of the observer $(t,~x,~y,~z)$ (unprimed coordinates). The observer is assumed to be static relative to the background.

\section{Second Post-Minkowskian Harmonic Metric for a Moving Kerr Black Hole}\label{secion_HarmonicMetric}
Let $\bm{e}_i~(i=1,2,3)$ denote the unit vector of three-dimensional rectangular coordinate system. We assume $m$ and
$\bm{J} (=J\bm{e}_3)$ stand for the mass and angular momentum of a Kerr black hole, respectively. Then the harmonic metric of Kerr black hole in the center of mass's rest frame can be expressed as~\cite{LJ2014,LJ20142}:
{\footnotesize\begin{eqnarray}
&&ds^{2}=-dX^{2}_{0}+\frac{R^2(R+m)^2+a^2X^2_3}{\left(R^2+\frac{a^2}{R^2}X^2_3\right)^2}\left[\frac{\left(\mathbf{X}\!\cdot\! d\mathbf{X}+\frac{a^2}{R^2}X_3dX_3\right)^{2}}{R^2+a^2-m^2} \!+\!\frac{X_3^2}{R^2} \frac{\left(\mathbf{X}\!\cdot\! d\mathbf{X}-\frac{R^2}{X_3} d X_3\right)^2}{R^2-X_3^2}\right]\nn\\
&&~~~~~+\frac{(R+m)^{2}+a^{2}}{R^2-X_3^2}\!\!\left[\!\frac{R^2m^2a \! \left(R^2\!-\!X_3^2\right) \! \left(\mathbf{X}\! \cdot \! d\mathbf{X}+\frac{a^2}{R^2}X_3dX_3\right)}{(R^2+a^2-m^2)(R^2+a^2)(R^4+a^2X^2_3)}\!+\!\frac{R\left(X_2dX_1\!-\!X_1dX_2\right)}{R^2+a^2}\!\right]^2 \nn \\
&&+\frac{2m(R\!+\!m)}{(R\!+\!m)^2\!+\!\frac{a^2}{R^2}X^2_3}\!\!\left[\!\frac{Rm^2a^2 \! \left(R^2\!-\!X_3^2\right) \!
\left(\mathbf{X}\!\cdot\!d\mathbf{X}+\frac{a^2}{R^2}X_3dX_3\right)}{(R^2+a^2-m^2)(R^2+a^2)\left(R^4+a^2X^2_3\right)}
\!+\!\frac{a\left(X_2dX_1\!-\!X_1dX_2\right)}{R^2+a^2}\!+\!dX_{0}\!\right]^2~,~~~~ \label{HarmonicKerr}
\end{eqnarray}}
where $\mathbf{X}\!\cdot\! d\mathbf{X}\equiv X_1dX_1\!+\!X_2dX_2\!+\!X_3dX_3$ and $a \equiv \frac{J}{m}$ is the angular momentum per mass. $R$ is related to $X_1,~X_2$ and $X_3$ via the relation $\frac{X_1^2+X_2^2}{R^2+a^2}+\frac{X_3^2}{R^2}=1$. Notice that here $X_{\mu}$ denotes the 4-dimensional coordinates $x'^{\mu}=(t',~x',~y',~z')$ for display convenience. There are many efforts contributed to derive the Kerr metric in harmonic coordinates, and Eq.~\eqref{HarmonicKerr} has been compared with the results of other researchers in Ref.~\cite{LJ2014}. If we drop all Kerr-related terms $(a=0)$, this equation will reduce to the Schwarzschild metric in the harmonic coordinates~\cite{Weinberg1972,Poisson2007}.

In the limit of weak field, the metric can be reduced to
\begin{eqnarray}
&&g_{00}=-1-2\Phi-2\Phi^2~, \label{g00s}\\
&&g_{0i}=\zeta_i~, \label{g0is}\\
&&g_{ij}=(1-\Phi)^2\delta_{ij}+\Phi^2\frac{X_iX_j}{R^2}~, \label{gijs}
\end{eqnarray}
where the order$~1/R^2$ is kept. $\Phi\equiv-\frac{m}{R}$ is Newtonian gravitational potential, and for given $X_1,~X_2,~X_3$,
\begin{eqnarray}
R=\sqrt{\frac{X_1^2+X_2^2+X_3^2-a^2+\sqrt{(X_1^2+X_2^2+X_3^2-a^2)^2+4a^2X_3^2}}{2}}~,
\end{eqnarray}
which reduces to $R=\sqrt{X_1^2+X_2^2+X_3^2}$ when the angular momentum vanishes $(a=0)$. $\bm{\zeta}\equiv \frac{2am}{R^3}\left(\bm{X} \times \bm{e_3}\right)$~denotes the vector potential due to the rotation of Kerr black hole. This approximate metric is in accord with the results of multipole expansions~\cite{Thorne1980}, as shown in Appendix A.

Using Lorentz transformation, one can obtain the metric for an arbitrarily moving Kerr black hole from Eqs.~\eqref{g00s}\,-\,\eqref{gijs}. Here we denote the translational velocity in arbitrary direction of this black hole as ${\bm v}=v_1\bm{e}_1+v_2\bm{e}_2+v_3\bm{e}_3~$. The general Lorentz transformation between $(t',~x',~y',~z')$ and the coordinate frame of the background $(t,~x,~y,~z)$ can be written as
\begin{equation}
x'^\alpha = \Lambda^\alpha_\beta x^\beta~, \label{lorentz0}
\end{equation}
and
\begin{eqnarray}
&&\Lambda_0^0=\gamma~,  \label{LorentzTran1} \\
&&\Lambda_0^i=\Lambda_i^0=-v_i\gamma~,  \label{LorentzTran2} \\
&&\Lambda^i_j=\delta_{ij}+v_iv_j\frac{\gamma-1}{v^2}~,  \label{LorentzTran4}
\end{eqnarray}
where $\gamma= (1-v^2)^{-\scriptstyle \frac{1}{2}}$ is Lorentz factor and $v^2=\bm{v}^2=v_1^2+v_2^2+v_3^2$. Following the definition of covariant metric tensor $g_{\mu\nu}=g'_{\rho\sigma}\Lambda^{\rho}_{\mu} \Lambda^{\sigma}_{\nu}$, we can write down the second post-Minkowskian metric of the moving Kerr black hole:
{\small\begin{align}
&g_{00}=-1-2\gamma^2\left(1+v^2\right)\Phi-\left(1+\gamma^2\right)\Phi^2-2\gamma^2\left(\bm{v}\cdot\bm{\zeta}\right)
+\frac{\gamma^2\Phi^2}{R^2}\left(\bm{v}\cdot\bm{X}\right)^2~,  \label{g00m}   \\
&g_{0i}=v_i\gamma^2\!\left(4\Phi\!+\!\Phi^2\right)
\!+\!\gamma\!\!\left[\zeta_i\!+\!\left(\!\frac{\gamma\!-\!1}{v^2}\!+\!\gamma\!\right)\!\left(\bm{v}\cdot\bm{\zeta}\right)\!v_i\!\right] \!-\!\frac{\gamma\Phi^2}{R^2}\!\!\left[\!\frac{v_i(\gamma\!-\!1)}{v^2}\left(\bm{v}\!\cdot\!\bm{X}\right)^2
\!+\!X_i\left(\bm{v}\!\cdot\!\bm{X}\right)\!\right]~,~\label{g0im}    \\
\nn&g_{ij}=\left(1-\Phi\right)^2\delta_{ij}-v_iv_j\gamma^2\left(4\Phi+\Phi^2\right)
-\gamma\left[\zeta_iv_j+\zeta_jv_i+\frac{2(\gamma-1)}{v^2}\left(\bm{v}\cdot\bm{\zeta}\right)v_iv_j\right]   \\
&~~~~~~~+\frac{\Phi^2}{R^2}\left[X_iX_j+\frac{v_i\,v_j\,(\gamma-1)^2}{v^4}\left(\bm{v}\cdot\bm{X}\right)^2
+\frac{(\gamma-1)\left(v_iX_j+v_jX_i\right)}{v^2}\left(\bm{v}\cdot\bm{X}\right)\right]~.~\label{gijm}
\end{align}}
For simplicity, here we formulate the metric in terms of the old coordinates, which are related to the new coordinates via $x'^{\alpha}= \Lambda^\alpha_\beta x^\beta$ as follows:
\begin{eqnarray}
&&x'=X_1=x+v_1\left[\frac{\gamma-1}{v^2}(\bm{v}\cdot\bm{x})-\gamma t\right]~,  \label{LT1} \\
&&y'=X_2=y+v_2\left[\frac{\gamma-1}{v^2}(\bm{v}\cdot\bm{x})-\gamma t\right]~,  \label{LT2} \\
&&z'=X_3=z+v_3\left[\frac{\gamma-1}{v^2}(\bm{v}\cdot\bm{x})-\gamma t\right]~.  \label{LT3}
\end{eqnarray}
In terms of the new coordinates, $R$, $\bm{\zeta}$, $\bm{v}\cdot\bm{X}$ and $\bm{v}\cdot\bm{\zeta}$ can also be written as
{\small\begin{eqnarray}
\nn&&R=\frac{1}{\sqrt{2}}\Big\{x^2+y^2+z^2\!-\!t^2+\gamma^2(\bm{v}\cdot\bm{x}\!-\!t)^2\!-\!a^2    \label{LT4} \\
&&~~+\sqrt{\big[x^2\!+\!y^2\!+\!z^2\!-\!t^2\!+\!\gamma^2(\bm{v}\cdot\bm{x}\!-\!t)^2\!-\!a^2\big]^2
\!+\!4a^2\big\{z\!+\!v_3\big[\frac{\gamma\!-\!1}{v^2}(\bm{v}\!\cdot\!\bm{x})\!-\!\gamma t\big]\big\}^2}~\Big\}^{\frac{1}{2}},  \label{LT5}  \\
&&\bm{\zeta}=\frac{2ma\left\{y\bm{e}_1-x\bm{e}_2+(v_2\bm{e}_1-v_1\bm{e}_2)\left[\frac{\gamma-1}{v^2}(\bm{v}\cdot\bm{x})-\gamma t\right]\right\}}
{R^{3}}~,  \label{LT6} \\
&&\bm{v}\cdot\bm{X}=\gamma\left(\bm{v}\cdot\bm{x}-v^2t\right)~,  \label{LT7}\\
&&\bm{v}\cdot\bm{\zeta}=\frac{2ma(v_1y-v_2x)}{R^{3}}~.  \label{LT8}
\end{eqnarray}}
Eqs.~\eqref{g00m} - \eqref{gijm} extend the metric presented in Ref.~\cite{HeLin2014b} to arbitrary direction of translational velocity of gravitational source.
In the limit of low velocity ($v\rightarrow 0$), Eqs.~\eqref{g00m} - \eqref{gijm} reduce to the results of post-Newtonian approximation~\cite{Weinberg1972}
\begin{eqnarray}
&&g_{00}=-1-2\left(1+2v^2\right)\Phi-2\Phi^2~,  \label{g00mPN}   \\
&&g_{0i}=4v_i\Phi\hspace*{2pt}+\hspace*{2pt}\zeta_i~,   \label{g0imPN}    \\
&&g_{ij}=\left(1-2\Phi\right)\delta_{ij}~,  \label{gijmPN}
\end{eqnarray}
where $\Phi=-m/\sqrt{x^2+y^2+z^2-t^2+\gamma^2(\bm{v}\cdot\bm{x}-t)^2}~$ up to the first post-Newtonian order.

\section{Compatibility between Lorentz Transformation Solution and Retarded Li\'{e}nard-Wiechert Potential}

Considering the characteristic of the linear gravitational perturbation theory, here we limit to discussing the first post-Minkowskian consistency with the solution of directly solving the linear field equations.

\subsection{Comparison with Mass-induced Li\'{e}nard-Wiechert Potential of a Moving Monopole}

\subsubsection{Retarded Li\'{e}nard-Wiechert Gravitational Potential}

For a massive point-like particle moving with the arbitrary constant speed $\bm{v}$ mentioned above, the corresponding energy-momentum tensor
in the background's rest frame can be written as \cite{Weinberg1972}
\begin{eqnarray}
&&T^{00}(\bm{x},t)=\gamma m\delta^3(\bm{x}-\bm{x}_0(t))~,  \label{T00} \\
&&T^{0i}(\bm{x},t)=v^i\gamma m\delta^3(\bm{x}-\bm{x}_0(t))~, \label{T0i} \\
&&T^{ij}(\bm{x},t)=v^iv^j\gamma m\delta^3(\bm{x}-\bm{x}_0(t))~, \label{Tij}
\end{eqnarray}
where $i,j=1,2$ or $3$ and $m$ denotes the static mass of the particle.

In the linear perturbation theory, the metric tensor $g_{\mu\nu}$ takes the form of
\begin{equation}
g_{\mu\nu}(\bm{x},t)=\eta_{\mu\nu}+h_{\mu\nu}(\bm{x},t)~,  \label{gij}
\end{equation}
where $\eta_{\mu\nu}=(-1,~1,~1,~1)$ denotes the Minkowski metric and the perturbation $h_{\mu\nu}(\bm{x},t)$ related to
Eqs. \eqref{T00} - \eqref{Tij} in the first post-Minkowskian approximation is \cite{KopeiSch1999}
\begin{equation}
h^{\mu\nu}(\bm{x},t)=h^{\mu\nu}_M(\bm{x},t)
=\frac{4\left(T^{\mu\nu}-\frac{1}{2}\eta^{\mu\nu}T^\lambda_\lambda\right)}{r(s)-\bm{v}(s)\cdot\bm{r}(s)}~.
\end{equation}
Here $s=s(t,\bm{x})$ represents the retarded time, and the symbol $M$ denotes the case that the perturbation
here only depends on the mass of the source. Explicitly, we have
\begin{eqnarray}
&&h_{00}(\bm{x},t)=\frac{2(1+v^2)\gamma m}{r(s)-\bm{v}(s)\cdot\bm{r}(s)}~, \label{h00} \\
&&h_{0i}(\bm{x},t)=-\frac{4v_i\gamma m}{r(s)-\bm{v}(s)\cdot\bm{r}(s)}~, \label{h0i} \\
&&h_{ij}(\bm{x},t)=\frac{4v_iv_j\gamma m+\frac{2m}{\gamma}\delta_{ij}}{r(s)-\bm{v}(s)\cdot\bm{r}(s)}~. \label{hij}
\end{eqnarray}
Notice that the retarded denominator $r(s)-\bm{v}(s)\cdot\bm{r}(s)$ for the uniform motion of gravitational source can be also expressed as~\cite{Oleg1997}
\begin{equation}
r(s)-\bm{v}(s)\cdot\bm{r}(s)\equiv\langle r-\bm{v}\cdot\bm{r}\rangle=r_0\sqrt{1-v^2\sin^2\theta}~, \label{lelation1}
\end{equation}
where $\theta$ is the angle between the present position vector $\bm{r}_0$ and the translational velocity vector $\bm{v}$, $r_0=|\bm{r}_0|$ and
the angle brackets denote the retardation symbol. Hence, Eqs. \eqref{h00} - \eqref{hij} become
\begin{eqnarray}
&&h_{00}(\bm{x},t)=\frac{2(1+v^2)\gamma m}{r_0\sqrt{1-v^2\sin^2\theta}}~, \label{h-00} \\
&&h_{0i}(\bm{x},t)=-\frac{4v_i\gamma m}{r_0\sqrt{1-v^2\sin^2\theta}}~, \label{h-0i} \\
&&h_{ij}(\bm{x},t)=\frac{4v_iv_j\gamma m+\frac{2m}{\gamma}\delta_{ij}}{r_0\sqrt{1-v^2\sin^2\theta}}~. \label{h-ij}
\end{eqnarray}
Therefore, Eq.~\eqref{gij} has been expressed in terms of the present time-dependent metric via quantities of present time for comparison.

\subsubsection{First Post-Minkowskian Metric via Coordinate Transformation} \label{firstpost-MCT}

\begin{figure*}
  \centering
  \includegraphics[width=12cm]{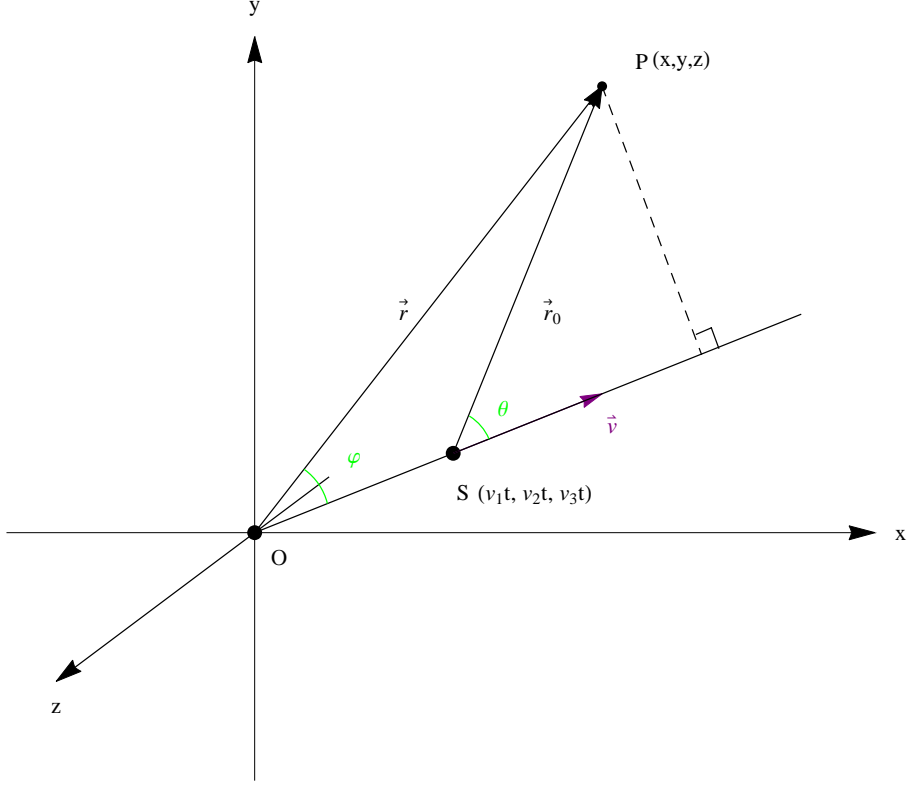}
  \caption{Sketch map of the geometrical relations between retarded position vector $\bm{r}$ and present position vector $\bm{r}_0$ for the moving Schwarzschild black hole. $P(x,y,z)$ and $S$ denote the field point and source point, respectively. We locate the source $S$ at the origin $O$ when $t=0$ so that the source is located at the position $(v_{1}t, v_{2}t, v_{3}t)$ at the present time $t(>0)$. $\theta$ denotes the angle between the present position vector $\bm{r}_{0}$ and velocity vector $\bm{v}$. $\varphi$ denotes the angle between the retarded position vector $\bm{r}$ and $\bm{v}$. }
  \label{figure1}
\end{figure*}
Now we turn our attention to the result to be analyzed based on Lorentz transformation. Here we don't distinguish the gravitational field of a Schwarzschild black hole from that of one static point mass $m$, as done in Ref.~\cite{HeLin2014}. Thus, according to
Eqs.~\eqref{g00m} - \eqref{gijm}, the first post-Minkowskian time-dependent metric of a moving point mass with the arbitrary constant speed $\bm{v}$ is
\begin{eqnarray}
&&g_{00}=-1-2(1+v^2)\gamma^2\Phi , \label{h00-2} \\
&&g_{0i}=4v_i\gamma^2\Phi , \label{h0i-2} \\
&&g_{ij}=(1-2\Phi)\delta_{ij}-4v_iv_j\gamma^2\Phi . \label{hij-2}
\end{eqnarray}
Notice that $\Phi$ in Eqs.~\eqref{h00-2} - \eqref{hij-2} is reduced to $\Phi=-m/\sqrt{X_1^2+X_2^2+X_3^2}$ in the first post-Minkowskian limit and that the metric here is denoted by quantities of co-moving frame $X_i\equiv (X_0, X_1, X_2, X_3)$ of the translationally moving Schwarzschild black hole.

For the convenience of comparison, we need to express $\Phi$ in terms of quantities of the background's rest frame. Using Lorentz transformation, Eq.~\eqref{lorentz0} between the co-moving frame and background's rest frame, we obtain $R$ in $\Phi$ as the function of the present distance $r_{0}$ between field point $P$ and source point $S$ as follow
\begin{eqnarray}
\nn&&R=\sqrt{\left(\Lambda^1_jx^j\right)^2+\left(\Lambda^2_jx^j\right)^2+\left(\Lambda^3_jx^j\right)^2}~    \\
&&\hspace*{11pt}=\gamma\sqrt{r_{0}^2-\left[v^2r^2-(\bm{v}\cdot\bm{x})^2\right]}~, \label{R}
\end{eqnarray}
where $r\equiv|\bm{r}|=\sqrt{x^2+y^2+z^2}$ denotes the distance between field point and the origin $O$ of coordinate frame $(t, x, y, z)$. Generally, the geometrical relations of the related quantities are showed in Fig. \ref{figure1}. For the velocity $\bm{v}$ in arbitrary direction, the relation below holds:
\begin{equation}
r_{0}\sin\theta=r\sin\varphi~.
\end{equation}

Therefore, Eq.~\eqref{R} is simplified to
\begin{equation}
R=\gamma r_{0}\sqrt{1-v^2\sin^2\theta}~. \label{R2}
\end{equation}
We rewrite Eqs.~\eqref{h00-2} - \eqref{hij-2} as follows:
\begin{eqnarray}
&&g_{00}=-1+\frac{2(1+v^2)\gamma m}{r_{0}\sqrt{1-v^2\sin^2\theta}} , \label{h00-3} \\
&&g_{0i}=-\frac{4v_i\gamma m}{r_{0}\sqrt{1-v^2\sin^2\theta}} , \label{h0i-3} \\
&&g_{ij}=\delta_{ij}+\frac{4v_iv_j\gamma m+\frac{2m}{\gamma}\delta_{ij}}{r_{0}\sqrt{1-v^2\sin^2\theta}}~, \label{hij-3}
\end{eqnarray}
which is the same as the result by solving linearized field equation. For the special case that the point mass moves along positive x-axis, namely $\bm{v}=v\bm{e}_1$, the exact solution~\cite{HeLin2014} via common Lorentz transformation is certainly consistent with the Li\'{e}nard-Wiechert Gravitational Potential. In the limit of low velocity (within post-Newtonian order), the gravielectric and gravimagnetic potentials from Eqs.~\eqref{h00-3} - \eqref{h0i-3} are consistent with that via solving gravitational wave equation in analogy to its electromagnetic counterpart~\cite{KopeiFoma2007}.

\subsection{Comparison with Spin-induced Li\'{e}nard-Wiechert Potential of a Moving Spinning Point Mass}
In this section, we compare the spin-induced Li\'{e}nard-Wiechert potential (see Eq.~(16) in Ref.~\cite{KopeiMash2002}) with the spin-induced terms in Eqs.~\eqref{g00m} - \eqref{gijm}
\begin{eqnarray}
&&g_{00}^S\sim-2\gamma^2\left(\bm{v}\cdot\bm{\zeta}\right)~, \label{Stermg00}    \\
&&g_{0i}^S\sim\gamma\!\left[\zeta_i+\left(\frac{\gamma-1}{v^2}+\gamma\right)\left(\bm{v}\cdot\bm{\zeta}\right)v_i\right]~, \label{Stermg0i}    \\
&&g_{ij}^S\sim-\gamma\left[\zeta_iv_j+\zeta_jv_i+\frac{2(\gamma-1)}{v^2}\left(\bm{v}\cdot\bm{\zeta}\right)v_iv_j\right]~. \label{Stermg33}
\end{eqnarray}

According to the results presented in Ref.~\cite{KopeiMash2002}, the spin-induced Li\'{e}nard-Wiechert potential reads
\begin{eqnarray}
h^{\alpha\beta}_S=\frac{4}{\gamma^2}\frac{r_\xi S^{\xi(\alpha} u^{\beta)}}{\left[r(s)-\bm{v}\cdot\bm{r}(s)\right]^3}~,  \label{LWhij}
\end{eqnarray}
which is determined by the energy-momentum tensor generated by the spin of the moving point mass. Here the symbol $S$ denotes the spin-dependence,
in contrast with the mass-dependence $(M)$ mentioned above.

The detailed comparison of Eqs.~\eqref{Stermg00} - \eqref{Stermg33} with Eq.~\eqref{LWhij} is performed in Appendix B, explicitly. We conclude that the spin-induced terms in our metric are also consistent with the retarded Li\'{e}nard-Wiechert potential solution, except for an additional factor $1/\gamma$ which is missing in Eq.~(15) in Ref.~\cite{KopeiMash2002} and has also been noticed in previous literatures~\cite{Brugmann2006,ZschSof2014}.

Since the linear metric perturbation $h^{\alpha\beta}$ for a moving spinning gravitational source can be divided into two parts, namely the mass-dependent part $h^{\alpha\beta}_M$ and the spin-induced part $h^{\alpha\beta}_S$ in the first post-Minkowskian approximation~\cite{KopeiMash2002}, the compatibility demonstrated above is helpful to verify the solution obtained by the method of Lorentz transformation.

\section{Equations of Motion of Photon and Massive Test Particle in Post-Newtonian Approximation} \label{secion_dynamics}
Based on Eqs.~\eqref{g00m} - \eqref{gijm}, we can derive the post-Newtonian equations of motion for test particles via calculating Christoffel symbols and then substituting them into the geodesic equations. We consider a massive test particle or a photon in the weak gravitational field of the moving Kerr black hole. Eqs.~\eqref{g00m} - \eqref{gijm} in the low-velocity limit reads
\begin{eqnarray}
&&g_{00}=-1-2(1+2v^2)\Phi-2\Phi^2-2\bm{v}\cdot\bm{\zeta}+O(\overline{v}^6)~,  \label{weak-g00mS} \\
&&g_{0i}=4v_i\Phi+\zeta_i+O(\overline{v}^5)~,  \label{weak-g0imS}    \\
&&g_{ij}=(1-2\Phi)\delta_{ij}+O(\overline{v}^4)~,  \label{weak-gijmS}
\end{eqnarray}
where $\overline{v}$ denotes typical velocity of a non-relativistic system in the post-Newtonian approximation.

Up to order of $\overline{v}^5/\overline{r}$ ($\overline{r}$ denotes typical separation of a system of particles), the equation of motion for a massive test particle can be written as
\small{
\begin{eqnarray}
&&\frac{d\bm{u}}{dt}\!=-\nabla\!\left(\!\Phi\!+\!2v^2\Phi\!+\!2\Phi^2\!+\!\bm{v}\!\cdot\!\bm{\zeta}\right)
\!-\!\frac{\partial \bm{\xi}}{\partial t}\!+\!\bm{u}\!\times\!\left(\nabla\!\times\!\bm{\xi}\right)
\!+\!3\bm{u}\frac{\partial \Phi}{\partial t}\!+\!4\bm{u}\left(\bm{u}\!\cdot\!\nabla\right)\Phi\!-\!\bm{u}^2\nabla\Phi~, ~~~~\label{d1}
\end{eqnarray}}
where $\bm{\xi}=4\bm{v}\Phi+\bm{\zeta}$. The equation of motion for photon, up to order $\overline{v}^3$, are obtained as
\small{
\begin{eqnarray}
&&\frac{d\bm{u}}{dt}=-(1+\bm{u}^2)\nabla\Phi+4(1-\bm{v}\cdot\bm{u})\bm{u}\left(\bm{u}\cdot\nabla\right)\Phi
+\bm{u}\times\left[\nabla\times\left(4\bm{v}\Phi\right)\right]+(3-\bm{u}^2)\bm{u}\frac{\partial \Phi}{\partial t}~. ~~~~\label{d2}
\end{eqnarray}}
One can notice that Eqs.~\eqref{d1} - \eqref{d2} are fully consistent with the first post-Newtonian equations of motion given in Ref.~\cite{Weinberg1972}.

Considering the fact that the angular momentum per mass $a$ is less than mass $m$, we can see that the effects of the terms with $\Phi^2$ may be larger than that of the term containing $\bm{\zeta}$. When $v = 0$, we have $\bm{\xi}=\bm{\zeta}$, and Eqs. \eqref{d1} and \eqref{d2} reduce to describe the post-Newtonian equations of motion of test particles in the gravitational field of Kerr black hole~\cite{LJ2014}. When $\bm{\zeta}=0$, Eqs. \eqref{d1} and \eqref{d2} reduce to the post-Newtonian equations of motion of test particles in field of a moving Schwarzschild black hole~\cite{WuckSper2004,HeLin2014}.

\section{Conclusion}\label{secConclusion}
We have applied a general Lorentz transformation to the harmonic Kerr metric, and obtained the second post-Minkowskian metric for a moving Kerr black hole with an arbitrary constant speed. The perturbation of the trajectory of test particles due to the terms with $\Phi^2$ in this metric may be larger than that of the term containing the source's spin $a$. We have also illustrated that the metric obtained via Lorentz transformation is in agreement with the Li\'{e}nard-Wiechert solution for a moving and spinning point mass in the first post-Minkowskian approximation, not limiting to low velocity. Furthermore, the resulting metric via Lorentz transformation doesn't depend on the choice of spin supplementary conditions while the Li\'{e}nard-Wiechert gravitational potential is based on it. As an application, we calculated the post-Newtonian equations of motion for the massive test particle and photon based on the metric, which can also be used to investigate the second-order gravitational deflection of light as well as massive test particles, in the field of a moving Kerr black hole.

\section*{Acknowledgments}
This work was supported in part by the Program for New Century Excellent Talents in University (No. NCET-10-0702), the National Basic Research Program of China (973 Program) Grant No. 2013CB328904, and the Ph.D. Programs Foundation of Ministry of Education of China (No. 20110184110016), as well as the Fundamental Research Funds for the Central Universities.

\appendix

\section{Comparison between Eqs. (2) - (4) with the multipole expansions}
The multipole expansions for static field in Ref.~\cite{Thorne1980} (see, Eqs. (8.13) and (10.6)) are given as
\begin{eqnarray}
&&g_{00}=-1+\frac{2g}{r}-\frac{2g^2}{r^2}+O(1/r^3)~, \label{MP1} \\
&&g_{0j}=-\frac{2\epsilon_{jpq}S_pn_q}{r^2}+O(1/r^3)~,  \label{MP2} \\
&&g_{ij}=\delta_{ij}\left(1+\frac{2g}{r}\right)+\frac{g^2}{r^2}(\delta_{ij}+n_in_j)+O(1/r^3)~, \label{MP3}
\end{eqnarray}
where $n_i=\frac{x_i}{r}$ and $\epsilon_{ijk}$ is the completely antisymmetric Minkowskian tensor with $\epsilon_{123}=+1$. $g(=M=m)$ and $S_i$ denote the constant mass and angular momentum of the source, respectively. For the case of $\bm{J}=ma\bm{e}_3$, we rewrite the component $g_{0i}$ as
\begin{eqnarray}
&&g_{01}=-\frac{2\epsilon_{1pq}S_pn_q}{r^2}=-\frac{2(\epsilon_{123}S_2n_3+\epsilon_{132}S_3n_2)}{r^2}=\frac{2max_2}{r^3}~, \label{MP4} \\
&&g_{02}=-\frac{2\epsilon_{2pq}S_pn_q}{r^2}=-\frac{2(\epsilon_{213}S_1n_3+\epsilon_{231}S_3n_1)}{r^2}=-\frac{2max_1}{r^3}~, \label{MP5} \\
&&g_{03}=-\frac{2\epsilon_{3pq}S_pn_q}{r^2}=0~. \label{MP6}
\end{eqnarray}
It can be viewed obviously that Eqs.~\eqref{MP1} - \eqref{MP6} are consistent with Eqs.~\eqref{g00s} - \eqref{gijs}.

\section{Comparison with the spin-induced Li\'{e}nard-Wiechert potential}

Considering the transformation Eqs.~\eqref{lorentz0} - \eqref{LorentzTran4} and Eq.~\eqref{R2}, we present the spin-induced terms $g_{\mu\nu}^S$ in Eqs.~\eqref{g00m} - \eqref{gijm} explicitly in terms of the quantities of the background's rest frame as
\begin{eqnarray}
&&g_{00}^S\sim\frac{4ma(v_{2}x-v_{1}y)}{\gamma\left(r_0\sqrt{1-v^2\sin^2\theta}\right)^3}~, \label{E-Stermg00} \\
&&g_{01}^S\sim\frac{2ma\left[y-v_2t+v_1(v_1y-v_2x)+\frac{\gamma v_3(v_2z-v_3y)}{\gamma+1}\right]}
{\gamma\left(r_0\sqrt{1-v^2\sin^2\theta}\right)^3}~, \label{E-Stermg01}    \\
&&g_{02}^S\sim\frac{2ma\left[-x+v_1t+v_2(v_1y-v_2x)-\frac{\gamma v_3(v_1z-v_3x)}{\gamma+1}\right]}
{\gamma\left(r_0\sqrt{1-v^2\sin^2\theta}\right)^3}~, \label{E-Stermg02}    \\
&&g_{03}^S\sim \frac{2mav_3(v_1y-v_2x)\left(2\gamma+1\right)}{\gamma(\gamma+1)\left(r_0\sqrt{1-v^2\sin^2\theta}\right)^3}~, \label{E-Stermg03}    \\
&&g_{11}^S\sim-\frac{4mav_1\left[y-v_2t+\frac{\gamma v_3(v_2z-v_3y)}{\gamma+1}\right]}{\gamma\left(r_0\sqrt{1-v^2\sin^2\theta}\right)^3}~, \label{E-Stermg11}    \\
&&g_{22}^S\sim\frac{4mav_2\left[x-v_1t+\frac{\gamma v_3(v_1z-v_3x)}{\gamma+1}\right]}{\gamma\left(r_0\sqrt{1-v^2\sin^2\theta}\right)^3}~, \label{E-Stermg22}    \\
&&g_{33}^S\sim-\frac{4mav_3^2(v_1y-v_2x)}{(\gamma+1)\left(r_0\sqrt{1-v^2\sin^2\theta}\right)^3}~, \label{E-Stermg33}  \\
&&g_{12}^S\sim\frac{2ma\left[\left(t-\frac{\gamma v_3z}{\gamma+1}\right)(v_2^2-v_1^2)
+(v_1x-v_2y)\left(1-\frac{\gamma v_3^2}{\gamma+1}\right)\right]}{\gamma\left(r_0\sqrt{1-v^2\sin^2\theta}\right)^3}~, \label{E-Stermg12} \\
&&g_{13}^S\sim\frac{2mav_3\left\{v_2t-y-\frac{\gamma}{\gamma+1}\left[v_2(v_3z-v_1x)+y(v_1^2-v_3^2)\right]\right\}}
{\gamma\left(r_0\sqrt{1-v^2\sin^2\theta}\right)^3}~, \label{E-Stermg13}    \\
&&g_{23}^S\sim\frac{2mav_3\left\{x-v_1t+\frac{\gamma}{\gamma+1}\left[v_1(v_3z-v_2y)+x(v_2^2-v_3^2)\right]\right\}}
{\gamma\left(r_0\sqrt{1-v^2\sin^2\theta}\right)^3}~. \label{E-Stermg23}
\end{eqnarray}

On the other hand, substituting Eqs. (D1) - (D2) in Ref.~\cite{KopeiMash2002} and Eq.~\eqref{lelation1} into Eq.~\eqref{LWhij}, we write the explicit components of the perturbation with $\bm{J}=ma\bm{e}_3$ as follow:
\begin{equation}
h_{00}^S=\frac{4ma(x_1v_2-x_2v_1)}{\left(r_0\sqrt{1-v^2\sin^2\theta}\right)^3}~,  \label{EXPLICIT-LW-h00}
\end{equation}
{\small
\begin{eqnarray}
\nn&&h_{0i}^S=\!-\frac{2\left\{\!r\!\left(\bm{v}\!\times\!\bm{J}\right)^i\!+\!ma(x_1v_2\!-\!x_2v_1)v_i
\!+\!(x_1\epsilon_{1ik}\!+\!x_2\epsilon_{2ik}\!+\!x_3\epsilon_{3ik})\!
\left[\!J_k\!-\!\frac{v_k\gamma}{\gamma+1}\left(\bm{v}\!\cdot\!\bm{J}\right)\!\right]\!\right\}}
{\left(r_0\sqrt{1-v^2\sin^2\theta}\right)^3}~, \\ \label{EXPLICIT-LW-h0i} \\
\nn&&h_{ij}^S=2\Big\{(x_1\epsilon_{1ik}\!+\!x_2\epsilon_{2ik}\!+\!x_3\epsilon_{3ik})J_kv_j
\!+\!(x_1\epsilon_{1jk}\!+\!x_2\epsilon_{2jk}\!+\!x_3\epsilon_{3jk})J_kv_i-\frac{\gamma}{\gamma+1}\times  \\
\nn&&\hspace*{24pt}\left.\left(\bm{v}\cdot\bm{J}\right)\left[(x_1\epsilon_{1ik}+x_2\epsilon_{2ik}+x_3\epsilon_{3ik})v_kv_j
+(x_1\epsilon_{1jk}+x_2\epsilon_{2jk}+x_3\epsilon_{3jk})v_kv_i\right]  \right.  \\
&&\hspace*{24pt}+r\!\left[\left(\bm{v}\!\times\!\bm{J}\right)^iv_j+\left(\bm{v}\!\times\!\bm{J}\right)^jv_i\right]
\Big\}/\left(r_0\sqrt{1-v^2\sin^2\theta}\right)^3~,~~~~~~~\label{EXPLICIT-LW-hij}
\end{eqnarray}}
where $i,~j,~k=1,~2$ or $3$, $r_\alpha=(-r,~\bm{r})=(-t,~x_1,~x_2,~x_3)$ and the symbol of spin-dependence $S$ has been shifted up for display convenience. Namely,
\begin{eqnarray}
&&h_{00}^S=\frac{4ma(v_{2}x_1-v_{1}x_2)}{\left(r_0\sqrt{1-v^2\sin^2\theta}\right)^3}~,  \label{E-LW-h00}  \\
&&h_{01}^S=\frac{2ma\left[x_2-v_2t+v_1(v_1x_2-v_2x_1)+\frac{\gamma v_3(v_2x_3-v_3x_2)}{\gamma+1}\right]}
{\left(r_0\sqrt{1-v^2\sin^2\theta}\right)^3}~,  \label{E-LW-h01}  \\
&&h_{02}^S=\frac{2ma\left[-x_1+v_1t+v_2(v_1x_2-v_2x_1)-\frac{\gamma v_3(v_1x_3-v_3x_1)}{\gamma+1}\right]}
{\left(r_0\sqrt{1-v^2\sin^2\theta}\right)^3}~,  \label{E-LW-h02}  \\
&&h_{03}^S=\frac{2mav_3(v_1x_2-v_2x_1)\left(2\gamma+1\right)}{(\gamma+1)\left(r_0\sqrt{1-v^2\sin^2\theta}\right)^3}~,  \label{E-LW-h03}  \\
&&h_{11}^S=\frac{4mav_1\left[v_2t-x_2+\frac{\gamma v_3(v_3x_2-v_2x_3)}{\gamma+1}\right]}{\left(r_0\sqrt{1-v^2\sin^2\theta}\right)^3}~,  \label{E-LW-h11}  \\
&&h_{22}^S=\frac{4mav_2\left[x_1-v_1t+\frac{\gamma v_3(v_1x_3-v_3x_1)}{\gamma+1}\right]}{\left(r_0\sqrt{1-v^2\sin^2\theta}\right)^3}~,  \label{E-LW-h22}  \\
&&h_{33}^S=\frac{4mav_3^2(v_2x_1-v_1x_2)\gamma}{(\gamma+1)\left(r_0\sqrt{1-v^2\sin^2\theta}\right)^3}~,  \label{E-LW-h33}  \\
&&h_{12}^S=\frac{2ma\left[\left(t-\frac{\gamma v_3x_3}{\gamma+1}\right)(v_2^2-v_1^2)
+(v_1x_1-v_2x_2)\left(1-\frac{\gamma v_3^2}{\gamma+1}\right)\right]}{\left(r_0\sqrt{1-v^2\sin^2\theta}\right)^3}~, \label{E-LW-h12}  \\
&&h_{13}^S=\frac{2mav_3\left\{v_2t-x_2-\frac{\gamma}{\gamma+1}\left[v_2(v_3x_3-v_1x_1)+x_2(v_1^2-v_3^2)\right]\right\}}
{\left(r_0\sqrt{1-v^2\sin^2\theta}\right)^3}~, \label{E-LW-h13}
\end{eqnarray}
\begin{eqnarray}
&&h_{23}^S=\frac{2mav_3\left\{x_1-v_1t+\frac{\gamma}{\gamma+1}\left[v_1(v_3x_3-v_2x_2)+x_1(v_2^2-v_3^2)\right]\right\}}
{\left(r_0\sqrt{1-v^2\sin^2\theta}\right)^3}~. \label{E-LW-h23}
\end{eqnarray}

We can see that Eqs.~\eqref{E-Stermg00} - \eqref{E-Stermg23} coincide with Eqs.~\eqref{E-LW-h00} - \eqref{E-LW-h23}, except for an additional factor $1/\gamma$.


\vspace*{20pt}

\begin{thebibliography}{60}

  \bibitem{KopeiSch1999} S. M. Kopeikin and G. Sch\"{a}fer, Phys. Rev. D 60, 124002 (1999).
  \bibitem{Klioner2003} S. A. Klioner, Astron. Astrophys. 404, 783 (2003).
  \bibitem{KlionerPeip2003} S. A. Klioner and M. Peip, Astron. Astrophys. 410, 1063 (2003).
  \bibitem{KopeiMash2002} S. M. Kopeikin and B. Mashhoon, Phys. Rev. D 65, 064025 (2002).
  \bibitem{Sereno2002b} M. Sereno, Phys. Lett. A 305, 7 (2002).
  \bibitem{Sereno2005b} M. Sereno, Mon. Not. R. Astron. Soc. 359, L19 (2005).
  \bibitem{Sereno2007} M. Sereno, Mon. Not. R. Astron. Soc. 380, 1023 (2007).
  \bibitem{Bonvin2008} C. Bonvin, Phys. Rev. D 78, 123530 (2008).
  \bibitem{WuckSper2004} O. Wucknitz and U. Sperhake, Phys. Rev. D 69, 063001 (2004).
  \bibitem{KPSV2007} S. M. Kopeikin {\it et al.}, Phys. Lett. A 367, 276 (2007).
  \bibitem{BAI2008} B. Bertotti, N. Ashby and L. Iess, Class. Quantum Grav. 25, 045013 (2008).
  \bibitem{Kopeikin2009} S. Kopeikin, Phys. Lett. A 373, 2605 (2009).
  \bibitem{KopeiMak2007} S. M. Kopeikin and V. V. Makarov, Phys. Rev. D 75, 062002 (2007).
  \bibitem{ZscKlio2011} S. Zschocke and S. A. Klioner, Class. Quantum Grav. 28, 015009 (2011).
  \bibitem{Zschocke2011} S. Zschocke, arXiv:1105.3117.
  \bibitem{Linet2008} L. Lindegren {\it et al.}, Proc. IAU Symposium 3(S248), 217 (2007).
  \bibitem{CudHer1979} K. M. Cudworth and G. Herbig, Astron. J. 84, 548 (1979).
  \bibitem{EckGen1997} A. Eckart and R. Genzel, Mon. Not. R. Astron. Soc. 284, 576 (1997).
  \bibitem{Jarosz1998} M. Jaroszy\'{n}ski, Acta Astronom. 48, 413 (1998).
  \bibitem{Albroetal1999} M. D. Albrow {\it et al.}, Astrophys. J. 512, 672 (1999).
  \bibitem{Novatietal2010} S. C. Novati {\it et al.}, Astrophys. J. 717, 987 (2010).
  \bibitem{KopeiMaka2006} S. M. Kopeikin and V. Makarov, Astron. J. 131, 1471 (2006).
  \bibitem{LJ2014} C. Jiang and W. Lin, Gen. Rel. Grav. 46, 1671 (2014).
  \bibitem{LJ20142} W. Lin and C. Jiang, Phys. Rev. D 89, 087502 (2014).
  \bibitem{Weinberg1972} S. Weinberg, \textit{Gravitation and Cosmology: Principles and Applications of the General Theory of Relativity} (Wiley, New York, 1972).
  \bibitem{Poisson2007} E. Poisson, \textit{Post-Newtonian theory for the common reader}, Lecture Notes (2007).
  \bibitem{Thorne1980} S. K. Thorne, Rev. Mod. Phys. 52, 299 (1980).
  \bibitem{HeLin2014b} G. He and W. Lin, Int. J. Mod. Phys. D 23, 1450031 (2014).
  \bibitem{Oleg1997} O. D. Jefimenko, \textit{Electromagnetic Retardation and Theory of Relativity} (Electret Scientific, Star City, 1997), pp. 52-55.
  \bibitem{HeLin2014} G. He and W. Lin, Commun. Theor. Phys. 61, 270 (2014).
  \bibitem{KopeiFoma2007} S. M. Kopeikin and E. B. Fomalont, Gen. Rel. Grav. 39, 1583 (2007).
  \bibitem{Brugmann2006} M. H. Br\"{u}gmann, PhD Thesis, Friedrich-Schiller Universit\"{a}t Jena (2006).
  \bibitem{ZschSof2014} S. Zschocke and M. H. Soffel, arXiv:1403.5438.


\end{thebibliography}
\end{document}